# Comparative Evaluation of Data Stream Indexing Models

Mahnoosh Kholghi and MohammadReza Keyvanpour

*Abstract*—In recent years, the management and processing of data streams has become a topic of active research in several fields of computer science such as, distributed systems, database systems, and data mining. A data stream can be thought of as a transient, continuously increasing sequence of data. In data streams' applications, because of online monitoring, answering to the user's queries should be time and space efficient. In this paper, we consider the special requirements of indexing to determine the performance of different techniques in data stream processing environments. Stream indexing has main differences with approaches in traditional databases. Also, we compare data stream indexing models analytically that can provide a suitable method for stream indexing.

## I. INTRODUCTION

Data stream and its different applications have attached more attention in several fields of computer science, such as distributed systems, database systems, and data mining. Data stream can be conceived as a continuous and changing sequence of data that continuously arrives at a system to be stored or processed [1], [2]. Data stream has many applications such as network monitoring, telecommunication systems, stock markets and sensor networks.

Data Streams have different challenges in many aspects, such as computational, storage, querying and mining. Specifically, issues related to storing and processing data streams in a continuous and multiple input settings, raise new sort of challenges [3]. In a traditional database management system, storing these input data cannot be performed in an easy and convenient way; in contrast, data streams must be processed in an online manner to guarantee updated responses and fast query answering with minimum delay. Developing data stream management is an open and active research field [4].

Continuous queries that run indefinitely, unless a query lifetime has been specified, fit naturally into the mold of data stream applications. Examples of these queries include monitoring a set of conditions or events to occur, detecting a certain trend in the underlying raw data, or in general discovering relations between various components of a large real time system. The kinds of queries that are of interest from an application point of view can be listed as follows [5]: (1) monitoring aggregates, (2) monitoring or finding patterns, and (3) detecting correlations. Each of these queries requires data management over some history of values and not just over the most recently reported values. For example in case of aggregate queries, the system monitors whether the current window aggregate deviates significantly from that aggregate in most time periods of the same size. In case of correlation queries, the self-similar nature of sensor measurements may be reflected as temporal correlations at some resolution over the course of the stream. Therefore, the system has to maintain historical data along with the current data in order to be able to answer these queries.

There have been proposed a few indexing models according to data stream requirements. There are few papers concerning stream indexing. In literature we can find some papers about sliding windows indexing over data streams [6], [7]. A new model has been presented by Shivakumar and Garcia-Molina that solves the sliding window problems [8]. We focus on a multi-resolution indexing architecture. The architecture enables the discovery of "interesting" behavior online, provides flexibility in user query definitions, and interconnects registered queries for real-time and in-depth analysis [5].

In the following sections, we focus on the requirements of data stream processing and indexing. The rest of this paper is structured as follow. Section II describes the applicability of relational data indexing techniques in data stream management systems. In Section III, we review data stream indexing methods and their characteristics and challenges. Section IV compares these methods with analytical criterions. Section V concludes the paper.

## II. THE REQUIREMENTS OF DATA STREAM INDEXING

There are various structures able to be used as an index. In relational database management system we can find such examples that we describe them shortly in Table I.

Indexing method analysis in environment of data stream processing needs to take into consideration different requirements [9]. Relations are indexed by keys; in case of sequences, an order is imposed. But timeline domain indexing in searching task is potentially helpful. As it's mentioned before, due to diverse requirements in data stream environments, it's not possible to directly use the general indexing structures in those environments and only their primitive principles are applied in defining a new structure; because there are other measures that must be considered.

A system that manages these infinite heterogeneous data streams must satisfy the following requirements [10]:
- Mechanism for rate control must be in place to deal with data streams that are nearly always too fast for any indexing system.

Manuscript received March 21, 2012; revised May 2, 2012.
Mahnoosh Kholghi is with the Department of Electronic, Computer and IT Islamic Azad University Qazvin, Iran and Young Researchers Club, Qazvin Branch, Islamic Azad University, Qazvin, Iran (e-mail: m.kholghi@qiau.ac.ir).
MohammadReza Keyvanpour is with the Department of Computer Engineering Alzahra University Tehran, Iran (e-mail: keyvanpour@alzahra.ac.ir).

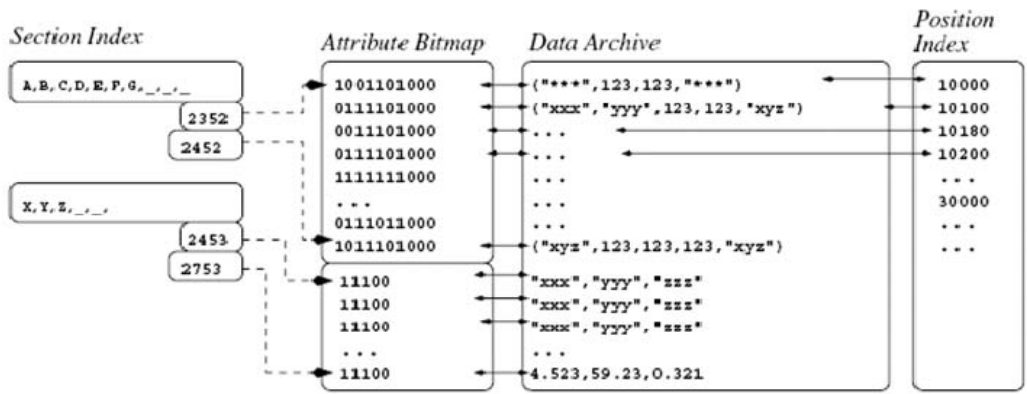

Fig. 1. The general structure of indexing files [10]

- There must be capacity control because the data streams are infinite while the storage space is finite.
- The structure of the data varies in time (i.e., heterogeneity of data streams), therefore adaptive indexing methods must be developed.
- Data should be efficiently stored.
- Volume can be huge, therefore everything must be appended-only; both indexing and query processing must only use forward-only file access.
- Users can query and filter the index streams efficiently.

TABLE I: DATA INDEXING TECHNIQUES

| Indexing methods in relational database management systems | Characteristics | Applicability in the structures of data stream indexing |
|---|---|---|
| Simple index | It is used on sorted files. Each record of index file contains two elements – key and pointer. | - |
| Indexes on unsorted files | The both elements of the pair are not in order. So, considering various queries, the set of index is needed. | - |
| B-tree | It has a tree-like structure ordering data blocks into tree structure. The tree is balanced if length of all paths is the same | A particular kind of it, $B^+$ tree, in time index [11]. |
| Hash index | Each pointer (address) of disk block containing a desired record is computed using a function (so called hash function) and the search key. Hash function maps the set of all search keys to the set of all records of blocks. | In STREAM system [12] |
| Bitmap index | It provides pointers to the rows in a table that contain a given key value. Each bit in the bitmap corresponds to a possible record. | In ArQss system [8] |

## III. DATA STREAM INDEXING MODELS

There are various indexing models for data steams:

### A. Bitmap Index Based Model

The problem of indexing continuous data streams in which data are heterogeneous in structure can be solved with bitmap indexing models. Such data streams arise naturally in many real-life scenarios such as sensor networks. This index structure uses bitmap based techniques to efficiently sketch the structures to allow space-efficient lossless archiving of the data stream. It also allows very fast query processing on the archived data stream.

Data stream indexing system, ArQSS (*Archiving and Querying Sensor Streams*) meets all the above requirements. ArQSS uses an adaptive bitmap based index structure to efficiently store incoming data readings into separate indexing files [10]. These files can be quickly accessed by ad-hoc user queries or stream filters. ArQSS offers a number of tuning parameters so it can be configured to performance optimization and controlled indexing rate. Several optimization techniques have been proposed to automatically configure ArQSS to achieve near-optimal performance, and a user-specified indexing rate. The general structure of indexing files is depicted in Fig. 1.

### B. Sliding Window Based Model

There are at least three reasons why sliding window indexes are useful [8]. The first is that application semantics require a sliding window. A second reason is that user interest in data may wane over time. A third reason is to reduce storage costs.

Sliding window indexes have been in use for many years; but the tremendous volumes of data that are today being generated in some applications makes it worthwhile to study these indexes carefully. Therefore, this method has two main problems [13]: storing on disk and updated on-line.

### C. Wave Indices

Several applications require indexing data of a past window of days. For this several techniques have been proposed to build wave indices [8]. In this method, the data of a new day can be efficiently added and old data can be quickly expired to maintain the required window. The main idea is to split the index into several parts so that deletions and insertions do not affect the entire index. Maintaining clustered order on disk as well as temporarily storing parts of the index in main memory is also discussed. The structure of this model is shown in Fig. 2.

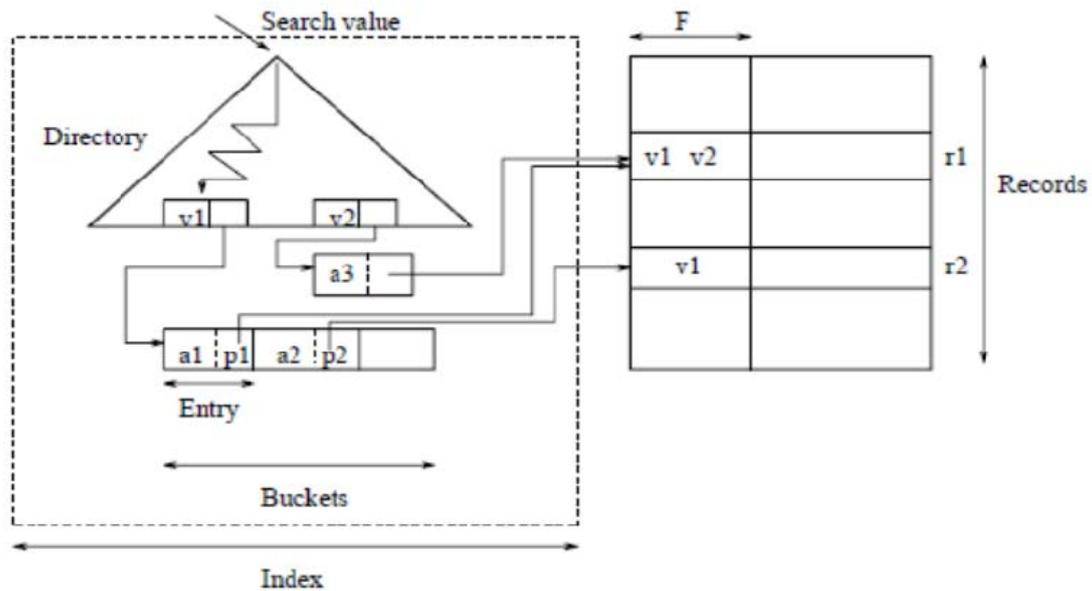

Fig. 2. Basic index structures [8]

Results indicate that each of wave indexing schemes has advantages and could be useful in some specific scenario, depending on what the central performance metrics are and on how much code can afford to write.

*D. Time Index Model*

Time index on checkpoints are a novel indexing technique for temporal data streams which are sorted on the effective start time attribute [11]. The index can be exploited a processing complex temporal pattern queries (such as multiway joins). In this approach indexes are built by periodically checkpoint the execution of query on along the time axis, and checkpoints are in turn indexed on their checkpoint times. This process is shown in Fig. 3.

Conventional methods such as $B^+$ tree can be used for implementing this type of indexing. For example, checkpoints are stored at leaf nodes of a $B^+$ tree as variable length records.

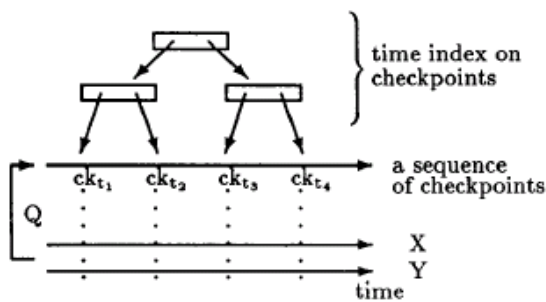

Fig. 3. Checkpointing a query execution and time index on checkpoints [11]

*E. Multi-resolution Indexing Model*

In this section, we introduce a multi-resolution indexing architecture. Multi-resolution approach imposes an inherent restriction on what constitutes a meaningful query. The core part of the scheme is the feature extraction at multiple resolutions. A dynamic index structure is used to index features for query efficiency [14].

The key architecture aspects are:
- The features at higher resolutions are computed using the features at lower resolutions; therefore, all features are computed in a single pass.
- The system guarantees the accuracy provided to user queries by provable error bounds.
- The index structure has tunable parameters to trade accuracy for speed and space. The per-item processing cost and the space overhead can be tuned according to the application requirements by varying the update rate and the number of coefficients maintained in the index structure.
- 

## IV. COMPARATIVE EVALUATION OF DATA STREAMS INDEXING MODELS

Stream indexing models can be evaluated in two distinct ways through requirements related measures. These evaluations are shown in Table II. Different data stream systems have their specific and diverse features and requirements. For evaluating indexing models that are applied in these systems we have assumed that there is no constant rate for input streams and consequently more data processing is needed to answer the queries.

Based on our comparison, multi-resolution indexing architecture model is time and space efficient and highly improve query answering in compare with other methods. Furthermore, this model considerably reduces the time unit which is needed for processing each data item and also minimizes the space needed for calculation. The structure of this indexing has acceptable time and space complexity. This complexity depends on updating rate and the number of coefficients conserving factors.

TABLE II: COMPARATIVE EVALUATION OF STREAM INDEXING MODELS BASED ON QUALITY MEASURES

|  | Storage Space | Online Updating | Proper Storage model |
|---|---|---|---|
| **Sliding Window Model** | Poor | Poor | Average |
| **Timeline domain indexing** | Average | Good | Poor |
| **Wave Indexing** | Good | Average | Average |
| **Bitmap Indexing Model** | Good | Average | Good |
| **Multi-resolution Indexing** | Good | Good | Good |

The timeline indexing models are mostly suitable for temporary data stream management systems and they have not proper performance to store data for a long time.

Based on evaluation measure, bitmap indexing model is an efficient model. When streams have no constant input rate and each tuples needs to be processed, sliding window model does not perform well in this condition. Wave indexing model has almost solve the problems of sliding window model but still has low performance for storing data in heterogeneous data stream systems whit continuous queries.

## V. CONCLUSIONS

This paper focuses on challenges and requirements of data stream indexing. Due to the nature of streaming data, indexing models for data streams are different from those models that are applied in relational database management systems. These differences are rooted in infinite, dynamic and in some cases heterogeneous nature of data streams. Each of data streams indexing models are used in a specific system and adapt to the requirements of those systems. Those models which are efficient based on different performance measures guaranty the query efficiency regarding to space and time issues. By fully review various indexing models, in this paper we presented a comparative evaluation of indexing models for data streams based on significant and well-known challenges in this field.

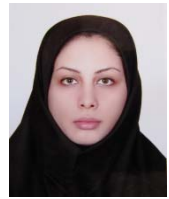

**Mahnoosh Kholghi** received her B.S. in Software Engineering from Islamic Azad University, Karaj Branch, Karaj, Iran. She also received her M.S. in Software Engineering at Islamic Azad University, Qazvin Branch, Qazvin, Iran. Her research interests include Data Stream Mining and Machine Learning.

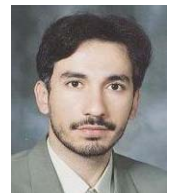

**MohammadReza Keyvanpour** is an Assistant Professor at Alzahra University, Tehran, Iran. He received his B.S. in Software Engineering from Iran University of Science &Technology, Tehran, Iran. He received his M.S. and Ph.D. in Software Engineering from Tarbiat Modares University, Tehran, Iran. His research interests include image retrieval and data mining.